\documentclass[pra,showpacs,superscriptaddress,twocolumn]{revtex4-1}

\usepackage[dvipdfmx]{graphicx}
\usepackage{amsmath}
\usepackage{mathtools}
 
\graphicspath{{graphics/}}  
\usepackage{color}
\usepackage{cancel}
\usepackage{ulem}
\usepackage{comment} 
\usepackage{bm} 
\usepackage{braket}
\usepackage{here}  

\newcommand{\LiNb}{\mathrm{LiNbO}_3}

\usepackage{float}

\begin{document}
\title{Quantitative evaluation method for magnetoelastic coupling between\\ surface acoustic waves and spin waves using electrical and optical measurements}

\author{Haruka~Komiyama}
\email{komiyama.haruka.64v@st.kyoto-u.ac.jp}
\affiliation{Institute for Chemical Research (ICR), Kyoto University, Gokasho, Uji, Kyoto 611-0011, Japan}

\author{Ryusuke~Hisatomi}
\email{hisatomi.ryusuke.2a@kyoto-u.ac.jp}
\affiliation{Institute for Chemical Research (ICR), Kyoto University, Gokasho, Uji, Kyoto 611-0011, Japan}
\affiliation{Center for Spintronics Research Network (CSRN), Kyoto University, Gokasho, Uji, Kyoto 611-0011, Japan}
\affiliation{PRESTO, Japan Science and Technology Agency, Kawaguchi, Saitama 332-0012, Japan}

\author{Kotaro~Taga}
\affiliation{Institute for Chemical Research (ICR), Kyoto University, Gokasho, Uji, Kyoto 611-0011, Japan}
\author{Hiroki~Matsumoto}
\affiliation{Institute for Chemical Research (ICR), Kyoto University, Gokasho, Uji, Kyoto 611-0011, Japan}
\author{Takahiro~Moriyama}
\affiliation{PRESTO, Japan Science and Technology Agency, Kawaguchi, Saitama 332-0012, Japan}
\affiliation{Department of Materials Physics, Nagoya University, Nagoya, Aichi 464-8603, Japan}
\author{Hideki~Narita}
\affiliation{Institute for Chemical Research (ICR), Kyoto University, Gokasho, Uji, Kyoto 611-0011, Japan}
\affiliation{PRESTO, Japan Science and Technology Agency, Kawaguchi, Saitama 332-0012, Japan}
\author{Shutaro~Karube}
\affiliation{Institute for Chemical Research (ICR), Kyoto University, Gokasho, Uji, Kyoto 611-0011, Japan}
\affiliation{Center for Spintronics Research Network (CSRN), Kyoto University, Gokasho, Uji, Kyoto 611-0011, Japan}
\affiliation{PRESTO, Japan Science and Technology Agency, Kawaguchi, Saitama 332-0012, Japan}
\author{Yoichi~Shiota}
\affiliation{Institute for Chemical Research (ICR), Kyoto University, Gokasho, Uji, Kyoto 611-0011, Japan}
\affiliation{Center for Spintronics Research Network (CSRN), Kyoto University, Gokasho, Uji, Kyoto 611-0011, Japan}
\author{Teruo~Ono}
\email{ono@scl.kyoto-u.ac.jp}
\affiliation{Institute for Chemical Research (ICR), Kyoto University, Gokasho, Uji, Kyoto 611-0011, Japan}
\affiliation{Center for Spintronics Research Network (CSRN), Kyoto University, Gokasho, Uji, Kyoto 611-0011, Japan}

\date{\today}

\begin{abstract}
{
Coupling and hybridization of different elementary excitations leads to new functionalities.
In phononics and spintronics, magnetoelastic coupling between Rayleigh-type surface acoustic wave (SAW) and spin wave (SW) has recently attracted much attention.
Quantitatively evaluating and comparing the coupled system are essential to develop the study of the magnetoelastic SAW-SW coupling.
So far, previous studies of SAW-SW coupling have employed a quantity called coupling strength. 
However, it is still challenging to compare the coupling strength values among studies fairly because the quantity depends on the device geometry and the applied magnetic field angle, which are not unified among the previous studies.
Here, we focus on a practical constant composed of a magnetoelastic constant and a strain amplitude that depends only on the material properties. 
We demonstrate a versatile evaluation technique to evaluate the practical constant by combining electrical measurements and optical imaging.
An essential part of the technique is an analysis that can be used under off-resonance conditions where SAW and SW resonance frequencies do not match.
Existing analysis can only handle the case under on-resonance conditions.
Our analysis makes it possible to observe the magnetoelastic couplings between SAW with resonance frequencies that can be imaged optically and SW with resonance frequencies in the gigahertz range.
Our demonstrated technique, which uses electrical and optical measurements under off-resonance conditions, can significantly advance research on SAW-SW coupled systems.
}
\end{abstract}

\pacs{
06.20.-f, 
74.25.Kc, 
75.30.Ds 
}

\maketitle

\section{Introduction}
Different elementary excitations localized at the same location may couple. Rayleigh-type surface acoustic wave (SAW) and spin wave (SW) are known to couple via magnetoelastic interactions in magnetic thin films on substrates~\cite{WD2011,DW2012,GM2015,LL2017,KH2021_LT,LZ2021,BT2021,YX2022}.
This coupled system has attracted considerable attention due to its potential as an information medium and functionality~\cite{SN2017,MK2022,PH2022,KA2022,RD2022,HA2023,SB2023,LP2023}.

It is essential to quantitatively evaluate and compare each coupled system to advance research on magnetoelastic SAW-SW coupling.
Previous studies used the coupling strength obtained from electrical measurements to evaluate the coupled system~\cite{HA2022,MY2024,HP2024}.
However, it is difficult to compare the coupling strength among different studies quantitatively because it depends on the spatial distribution of the SAW, the magnetic field angle, and the size of the magnetic film.
Here, we developed a method to evaluate a more practical constant that characterizes the coupled system by combining electrical and optical measurements. 
The practical constant, $bD$, is the quantity defined by the product of the magnetoelastic constant~$b$ and strain amplitude~$D$, and it depends mainly on the material.
Furthermore, we developed a method to separate the constant $bD$ into two types, longitudinal strain- and shear strain-derived.

SAW optical imaging is required to evaluate the constant $bD$.
Accurate imaging can be performed when the optical spot size is small relative to the wavelength of the SAW~\cite{HT2023,TH2021}.
For SAW on LN substrates, which is used in many related studies, the upper limit of the SAW frequency at which accurate optical imaging is possible is about $1\,\mathrm{GHz}$.
When using analytical methods of the magnetoelastic coupling under on-resonance conditions, where the resonance frequencies of SAW and SW match, as used in previous studies~\cite{WD2011,DW2012}, the evaluation of $bD$ is possible when SAW and SW frequencies are the same and below $1\,\mathrm{GHz}$.
In general, the resonance frequency of SW in ferromagnets is in the order of gigahertz. 
Thus, the number of magnetic materials that support SW below $1\,\mathrm{GHz}$ is limited.
To overcome this limitation, we developed an analytical method to detect the magnetoelastic coupling even under off-resonance conditions.
We demonstrated that the practical constant $bD$ can be evaluated by detecting the magnetoelastic coupling between a SAW with a frequency of $\sim800\,\mathrm{MHz}$ and a SW with a frequency of several gigahertz and then using the SAW's optical imaging results.

Chapter 2 briefly describes a theoretical model of the SAW-SW coupling system. Then, in Chapter 3, we characterize the SAW using optical measurements and evaluate the magnetoelastic coupling strength using electrical measurements. 
Finally, the practical constants are estimated by combining these results.

\begin{figure}[t]
\begin{center}
\includegraphics[width=8.6cm,angle=0]{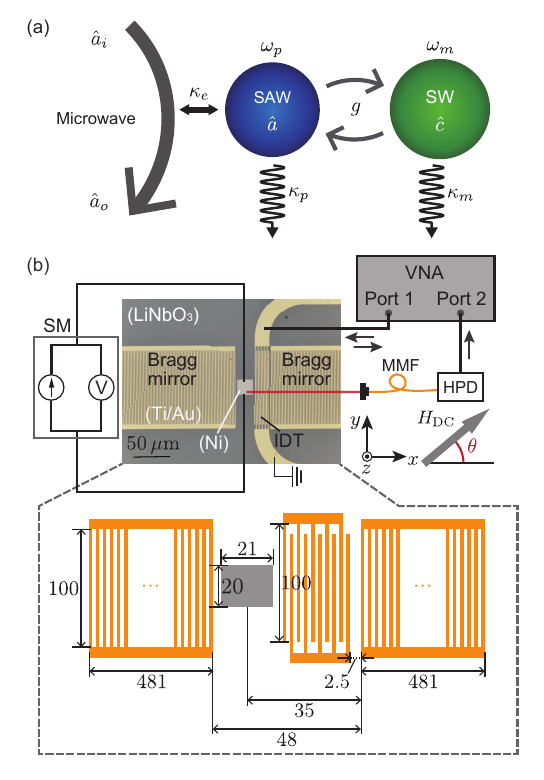}
\caption{ 
(a)~Architecture of the SAW-SW coupled system.
The system consists of two coupled harmonic oscillator modes, a SAW resonator mode $\hat{a}$, whose energy is specified by $\hbar \omega_{p}$, and spin wave mode $\hat{c}$, whose energy is specified by $\hbar \omega_{m}$, and these are coupled at a rate $g$.
An input (output) itinerant microwave field mode $\hat{a}_{i}$ ($\hat{a}_{o}$) is coupled to the coupled system through the SAW resonator mode at a rate $\kappa_{e}$.
$\kappa_{p}$ and $\kappa_{m}$ are rates of the intrinsic energy dissipation for the SAW resonator mode and the SW, respectively. 
(b)~Microscope image and design of a Fabry-Perot-type SAW resonator and a Ni film in Device~1 with a coordinate system.
The SAW resonator consists of a $0.5$-mm-thick YZ-cut $\LiNb$ substrate, Bragg mirrors, and interdigitated transducer (IDT).
Bragg mirrors and IDT are made by depositing $5\,\mathrm{nm}$ of Ti and $80\,\mathrm{nm}$ of Au. 
Two Bragg mirrors composed of $240$ fingers of $100$-$\mathrm{\mu m}$-long with a $1$-$\mathrm{\mu m}$ line and space are placed at a distance of $48\,\mathrm{\mu m}$. 
The IDT composed of ten fingers of $100$-$\mathrm{\mu m}$-long with a $1$-$\mathrm{\mu m}$ line and space are placed $2.5\,\mathrm{\mu m}$ apart from the edge of the Bragg mirror.
A Ni($50\,\mathrm{nm}$) film with a length of $21\,\mathrm{\mu m}$ and a width of $20\,\mathrm{\mu m}$ is deposited at the center between Bragg mirror and IDT.
The distance between the center of the film and the right Bragg mirror is $35\,\mathrm{\mu m}$.
A magnetic field $H_\mathrm{DC}$ is applied at an angle $\theta$ in the magnetic film plane.
A magnetoresistance in the Ni film is measured by a sourcemeter (SM).
The SAW resonator is electrically characterized by microwave reflection measurement using Port~1 of a vector network analyzer (VNA) through IDT.
Optical imaging of the SAW resonator is performed using light with a wavelength of $660\,\mathrm{nm}$ focused at the surface.
The light is incident in the $z$-axis direction and the optical path modulation added to the reflected light in the coaxial direction is observed.
The reflected light is received by a multi-mode fiber (MMF) and analyzed by a high-speed photodetector (HPD) connected to Port~2 of the VNA.}
\label{fig:coupledsystem+setup}
\end{center}
\end{figure}

\section{Theoretical Model of\\ coupled system} \label{sec:model}
In this section, we model a system in which the SAW and SW are coupled via magnetoelastic coupling in a magnetic thin film placed in a Fabry-Perot-type SAW resonator.
We then deduce the electrical reflection coefficient's model function, the magnetoelastic coupling strength that has been evaluated in previous studies~\cite{HA2022,MY2024,HP2024}, and the practical constants we focus on in this paper.

Figure~\ref{fig:coupledsystem+setup}(a) shows the architecture of a system in which a single SAW mode~$\hat{a}$ is coupled to a single SW mode~$\hat{c}$ and an itinerant microwave mode~$\hat{a}_{i}$.
We model the SAW, SW, and microwave coupling in a device consisting of a SAW resonator, a magnetic thin film, an interdigital transducer (IDT) shown in Fig.~\ref{fig:coupledsystem+setup}(b) with the architecture in Fig.~\ref{fig:coupledsystem+setup}(a).
The system is build on the two coupling Hamiltonians $H_\mathrm{me}$ and $H_\mathrm{p}$.
Here, $H_{\mathrm{me}}$ is the coupling between the SAW resonator mode $\hat{a}$ and the SW mode $\hat{c}$, given by
\begin{equation}
H_{\mathrm{me}}=\hbar g\left(\hat{a}^{\dagger}\hat{c}+\hat{c}^{\dagger}\hat{a}\right),
\label{eq:Hme_quant}
\end{equation}
with a coupling rate $g$.
$H_{\mathrm{p}}$ describes the coupling between an itinerant microwave mode $\hat{a}_{i}(\omega)$ and the SAW resonator mode $\hat{a}$, given by
\begin{equation}
H_\mathrm{p} = -i\hbar \sqrt{\kappa_{e}}\int_{-\infty}^{\infty}\frac{d\omega}{2\pi}\left(\hat{a}^{\dagger}\hat{a}_{i}(\omega) -\hat{a}_{i}^{\dagger}(\omega)\hat{a}\right),
\label{eq:Hp_quant}
\end{equation}
with a coupling rate $\kappa_{e}$.

\subsection{Electrical reflection coefficient $S_{11}$}
The total interaction Hamiltonian $H_\mathrm{t}=H_\mathrm{me}+H_\mathrm{p}$ with the intrinsic dissipations represented by rates $\kappa_{p}$ and~$\kappa_{m}$, for the SAW mode and the SW mode, respectively, defines the dynamics of the variables in the coupled system. 
For the SAW mode operator $\hat{a}$ we have the following Fourier-domain relation from the equation of motion:
\begin{equation}
\hat{a}(\omega) = \chi_{p}(\omega)\left(-\sqrt{\kappa_{e}}\hat{a}_{i}(\omega) -i(g)\hat{c}(\omega)\right),
\label{eq:a_motion}
\end{equation}
where the susceptibility $\chi_{p}(\omega)$ is defined as
\begin{equation}
\chi_{p}(\omega)=\left(-i\left(\omega-\omega_p\right) +\frac{\kappa_{p}+\kappa_{e}}{2}\right) ^{-1}.
\label{eq:kai_p}
\end{equation}
Here and hereafter the thermal and quantum noise terms are omitted. For the SW mode operator $\hat{c}$ we have
\begin{equation}
    \hat{c}(\omega) = \chi_{m}(\omega) \left(-g\hat{a}(\omega)\right),
    \label{eq:c_motion}
\end{equation}
where the susceptibility $\chi_{m}(\omega)$ is defined as
\begin{equation}
\chi_{m}(\omega) = \left( -i\left(\omega-\omega_{m}\right)+\frac{\kappa_{m}}{2}\right)^{-1}.
\label{eq:kai_m}
\end{equation}

 Solving the algebraic Eqs.~(\ref{eq:Hme_quant}) and (\ref{eq:Hp_quant}) with the boundary condition $\hat{a}_{o}(\omega)=\hat{a}_{i}(\omega) +\sqrt{\kappa_{e}}\hat{a}(\omega)$~\cite{CD2010}, the electrical reflection coefficient can be obtained as 
 \begin{eqnarray}
     S_{11}(\omega)&=&\left\langle\frac{\hat{a}_{o}(\omega)}{\hat{a}_{i}(\omega)}\right\rangle\\
     &=&\frac{i(\omega-\omega_p)-\frac{\kappa_p-\kappa_e}{2}+\frac{g^2}{i(\omega-\omega_m)-\kappa_m/2}}{i(\omega-\omega_p)-\frac{\kappa_p+\kappa_e}{2}+\frac{g^2}{i(\omega-\omega_m)-\kappa_m/2}}. \label{eq:S11model} 
 \end{eqnarray}

\subsection{Magnetoelastic coupling strength $g$ and practical constant $bD$}\label{subsec:g}
The classical magnetoelastic Hamiltonian $H_\mathrm{me}$ can be given by~\cite{K1958}
\begin{equation}
   H_\mathrm{me} = \int d^3 r \sum_{i,j,k,l=x,y,z} {B_{i,j,k,l}} \frac{m_i m_j}{M_S^2} e_{kl},\label{eq:Hme_classic1}
\end{equation}
where, $B_{i,j,k,l}$ is the magnetoelastic constant, $m_i$ is the component of magnetization, $M_S$ is the saturation magnetization of magnetic material, and $e_{kl}$ is the component of the strain tensor.
Since only the longitudinal strain $e_{xx}$ and shear strain $e_{zx}$ contributions are finite for Rayleigh-type SAW in in-plane magnetized films, Eq.~(\ref{eq:Hme_classic1}) becomes
\begin{equation}
   H_\mathrm{me} = \int d^3 r \left[b_1 \frac{m_x^2}{M_{{S}}^2} e_{xx} +  4\frac{b_2}{M_{{S}}^2} m_z m_x e_{zx}\right],\label{eq:Hme_classic2}
\end{equation}
where, $b_1=B_{x,x,x,x}$ and $b_2=B_{z,x,z,x}$.
The first and second terms represent the magnetoelastic coupling derived from longitudinal strain and shear strain, respectively.
As described in Appendix~\ref{Ap:g}, after extracting the time-dependent term of magnetization from Eq.~(\ref{eq:Hme_classic2}) and quantizing it, we obtain Eq.~(\ref{eq:Hme_quant}) where $g$ is expressed as
\begin{align}
    g &= \sqrt{\frac{\gamma k V_\mathrm{SW}}{\rho v_p M_{\rm{S}}V_\mathrm{SAW}}} \sqrt{(b_1 D_l \sin{2\phi})^2+(2b_2 D_s \cos{\phi})^2} \label{eq:g_allparameter} \\
    &=\sqrt{(g^c_l\sin{2\phi})^2+(g^c_s\cos{\phi})^2}\label{eq:g_forfit},
\end{align}
where
\begin{equation}
    g^c_l=b_1 D_l \sqrt{\frac{\gamma k V_\mathrm{SW}}{\rho v_p M_{\rm{S}}V_\mathrm{SAW}}} \label{eq:g_longi}   
\end{equation}
and
\begin{equation}
    g^c_s=2b_2D_s\sqrt{\frac{\gamma k V_\mathrm{SW}}{\rho v_p M_{\rm{S}}V_\mathrm{SAW}}}\label{eq:g_shear}.   
\end{equation}
Here, $\gamma$ is the gyromagnetic ratio of magnetic material, $k$ is the wavenumber of the SAW, $\rho$ is the mass density of the substrate, $v_p$ is the velocity of the SAW, $V_{\mathrm{SAW}}$ is the volume of the SAW, $V_{\mathrm{SW}}$ is the volume of the magnetic thin film, and $\phi$ is the relative angle of the magnetic field with respect to the direction of propagation of the SAW.
$g^c_l$ ($g^c_s$) is the longitudinal (shear) strain-derived magnetoelastic coupling strength, and $D_l$ ($D_s$) is the dimensionless quantity representing the amplitude of the longitudinal (shear) strain in the quantized displacement field. 
The relations between the strains $e_{xx}$ and~$e_{zx}$ in the classical displacement field and $D_l$ and $D_s$ are shown in Eqs.~(\ref{eq:exx_standingwave}) and (\ref{eq:ezx_standingwave}), respectively.

Previous studies mainly evaluated the coupling strength~$g$ in Eq.~(\ref{eq:g_allparameter})~\cite{HA2022,HP2024,MY2024}, which includes many parameters.
To enable a fair and accurate comparison of the magnetoelastic SAW-SW coupling across different material combinations, a normalization is crucial. 
The $b_1D_l$ and $b_2D_s$ in Eqs.~(\ref{eq:g_longi}) and (\ref{eq:g_shear}) are the most suitable and practical quantities to characterize the magnetoelastic SAW-SW coupling in each system because they depend only on the material.
By determining other values in Eqs.~(\ref{eq:g_longi}) and (\ref{eq:g_shear}) from literature, design, electrical measurements, and  optical imaging results, we can evaluate the constants, $b_1D_l$ and $b_2D_s$.
Note that attempts to estimate $b_{1,2}$ were made in previous studies with assumptions~\cite{HP2024,KH2020,KH2021_Bilayers}.
Also, note that $g^c_s$ is suggested to be mixed with an effect of the magneto-rotational coupling, which has the same angular dependence~\cite{YX2022,KH2020,MT1976,XY2020}.

\section{Experiments} \label{sec:experiments}

\subsection{Device preparation}\label{subsec:sample}
Figure~\ref{fig:coupledsystem+setup}(b) shows a microscope image of a SAW-SW device.
The part that appears to be gold is a Fabry-Perot-type SAW resonator consisting of two Bragg mirrors and interdigitated transducer (IDT) composed of Ti(5) and Au(80) by magnetron sputtering on a $0.5$-$\mathrm{mm}$-thick YZ-cut $\LiNb$ substrate, where the values in parentheses are the thicknesses in nanometers.
The lower panel in Fig.~\ref{fig:coupledsystem+setup}(b) shows the design drawing of the SAW-SW device.
The SAW resonator is fabricated so that the SAW propagation direction is along the crystalline $Z$-axis of the $\LiNb$ monocrystal.
The SAW resonator mode is coherently excited by driving the IDT with a microwave.
In the resonator's center, we deposit a rectangular Ni(50) film by magnetron sputtering.

To investigate the device geometry dependence of the magnetoelastic coupling, we fabricate two distinct SAW-SW devices, Device~1 and Device~2. 
This approach allows us to compare and analyze the effects of different device geometries on the coupling strength, $g^c_l$ and $g^c_s$ in Eqs.~(\ref{eq:g_longi}) and (\ref{eq:g_shear}).
The length in the $y$-direction of each Ni film is $20\,\mathrm{\mu m}$ in common, and the length in the $x$-direction is $21\,\mathrm{\mu m}$ for Device~1 and $13\,\mathrm{\mu m}$ for Device~2.
The center position of the Ni film relative to the SAW resonator is the same for the two devices.
Note that although the only difference in design is the Ni film size, in practice, the spatial distribution of SAW resonator modes can differ from device to device.
As described in Sec.\ref{OI}, the actual SAW mode volumes are different.

\subsection{Optical imaging}\label{OI}
To determine the SAW resonator modes in both devices, we first acquire two-dimensional plots of the amplitude and phase of the surface slope associated with the SAW.
The optical path modulation method, which we reported~\cite{HT2023,TH2021}, is employed as the measurement technique.
The method diagnoses SAW using path modulation of reflected light caused by the surface slope associated with SAW.
As shown in Fig. 1(b), light with a wavelength of $660\,\mathrm{nm}$ is launched down onto the device surface along the $z$ direction and focused to a diameter of about $1\,\mathrm{\mu m}$ at the surface.
The reflected light is input to a high-speed photodetector (HPD), and then the electrical signal from the HPD is input to Port~2 of a vector network analyzer (VNA).
The microwave from Port~1 of the VNA excites the SAW through the IDT, and the excited SAW coherently modulates the path of the reflected light.
By acquiring the transmitted signal $S_{21}$ while scanning the position of the optical spot, the SAW can be imaged.

\begin{figure}[t]
\begin{center}
\includegraphics[width=8.6cm,angle=0]{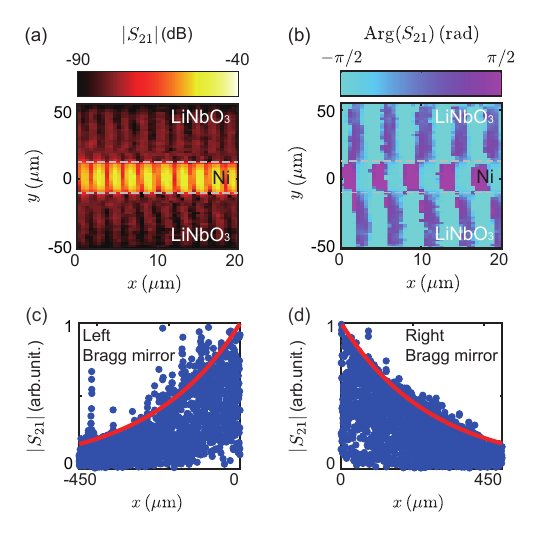}
\caption{
Imaging results of the SAW resonator in Device~1 by optical path modulation measurement.
Color map of (a) amplitude and (b) phase acquired by the VNA at a frequency $\omega/2\pi=803\,\mathrm{MHz}$.
(c)(d)~Position dependence of optical signals in the Bragg mirror.
The $x=0$ position represents the positon of the opposite ends of the Bragg mirrors.
The blue dots represent the optical signal within Bragg mirror on the (c) left side and (d) right side.
The red curves represent the results of exponential fittings of the envelopes.
Note that here the SAW is excited through the IDT connected to Port~1 of the VNA.
}
\label{fig:SAWimaging_Device1}
\end{center}
\end{figure}

Figures~\ref{fig:SAWimaging_Device1}(a) and \ref{fig:SAWimaging_Device1}(b) show the two-dimensional plots of the amplitude and the phase of the optical path modulation signal while the optical spot position is scanned in the $x$ axis every $0.5\,\mathrm{\mu m}$ and in the $y$ axis every $2\,\mathrm{\mu m}$ in a region ($20\,\mathrm{\mu m}\times 100\,\mathrm{\mu m}$) around center of the Ni film in Device~1.
The interval between the peak position of $|S_{21}|$ in Fig.~\ref{fig:SAWimaging_Device1}(a) is about $2\,\mathrm{\mu m}$, half the wavelength~$\lambda_{\mathrm{SAW}}$ of the excited SAW, and the phase in Fig.~\ref{fig:SAWimaging_Device1}(b) changes by exactly $\pi$ at each peak.
As expected, these results indicate that the SAW is a standing wave in the SAW resonator.
Furthermore, in the Ni film region in Figs.~2(a) and 2(b), there are no nodes in the $y$ direction in both phase and amplitude.
The result suggests that the fundamental SAW mode is excited in the Ni region.
In the same way, the excited SAW mode in Device~2 is also confirmed to be the fundamental mode (see Appendix C for details).
Note that, in contrast to the Ni region, the signal in the $\LiNb$ region may include intensity modulation due to photoelastic effect~\cite{SM2010}.
It is not easy to discuss the cause of the phase shift between the Ni and $\LiNb$ regions here.

We then experimentally determine the mode volume of the SAW using optical imaging.
The effective mode volume is defined as $V_\mathrm{SAW}=w\times l\times d$, where $w$ is the width of the IDT, $l$ is the effective length in the SAW propagation direction, and $d$ is the effective depth of the SAW resonator mode.
Since the SAW is localized on the order of SAW wavelength in the depth direction, we assume $ d\sim \lambda_\mathrm{SAW}$.
To experimentally determine the effective length $l$, we perform optical imaging of the SAW seeping into the Bragg mirror at $y=0\,\mathrm{\mu m}$ in Fig.~\ref{fig:SAWimaging_Device1}(a).
Figures~\ref{fig:SAWimaging_Device1}(c) and \ref{fig:SAWimaging_Device1}(d) show the $x$-position dependence of the optical path modulations signal.
The position $x=0$ is the opposite end of the two Bragg mirrors.
The blue dots represent the measured $S_{21}$ amplitudes, and the red lines represent the exponential fitting results.
The result show that the SAW seeps about $260\,\mathrm{\mu m}$ ($270\,\mathrm{\mu m}$) into the left (right) Bragg mirror, namely, $l=48+260+270=578\,\mathrm{\mu m}$.
For the SAW resonator in Device~1 we have $V_\mathrm{SAW}=2.2\times10^{-15}\,\mathrm{m^3}$.
Similarly, for the SAW resonator in Device~2 $V_\mathrm{SAW}=1.7\times10^{-15}\,\mathrm{m^3}$ (see Appendix C for details).

\subsection{Electrical measurements}
\subsubsection{Magnetic domain}
To determine how much magnetic field to apply to the Ni film to make it single magnetic domain, we measure the magnetic field dependence of the resistivity change due to the anisotropic magnetoresistance effect using the experimental setup in Fig.~\ref{fig:coupledsystem+setup}(b).
Figure~3(a) shows the magnetoresistance curve under the magnetic field direction of $\theta=90^{\circ}$.
The results of magnetoresistance curve measured at various $\theta$ suggest that the Ni film is a single magnetic domain 
when $\mu_0 H_\mathrm{DC}=10\,\mathrm{mT}$ or more (blue background) is applied at an arbitrary angle~$\theta$.
In a single magnetic domain, we can assume that a SW mode coupled to the SAW resonator mode has the same wavenumber as the SAW resonator mode.

\subsubsection{SW resonance frequency}

\begin{figure}[t]
\begin{center}
\includegraphics[width=8.6cm,angle=0]{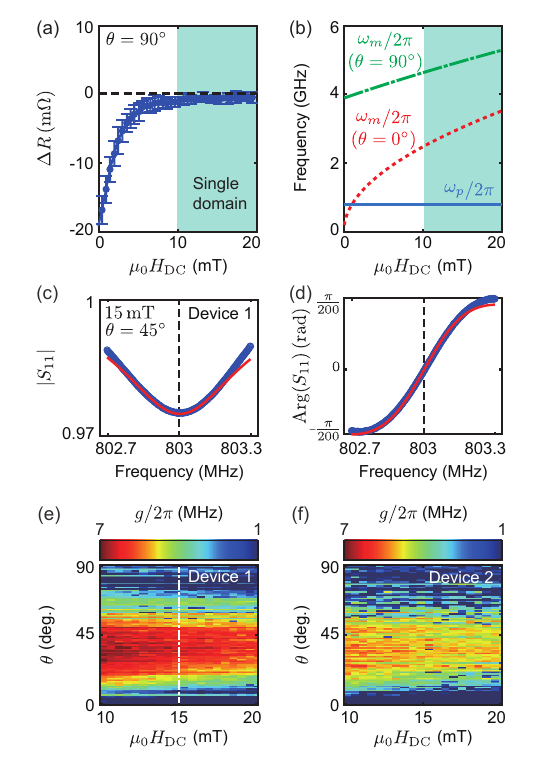}
\caption{
(a)~Anisotropic magnetoresistance measurement on Device~1 in $\theta=90^{\circ}$.
$\Delta R$ is obtained by subtracting the resistance for each magnetic field from the resistance at $\mu_0 H_{\mathrm{DC}}=200\,\mathrm{mT}$.
(b)~Magnetic field dependence of the resonance frequency of SW with the same wavenumber as SAW with a resonance frequency of $803\,\mathrm{MHz}$.
Green and red lines represent the resonance frequencies of SW at magnetic field angles $\theta=90^{\circ}$ and $\theta=0^{\circ}$, respectively.
(c)~Amplitude and (d)~phase spectra (blue dots) of the SAW resonator mode acquired by microwave reflection measurement $S_{11}$ at Port~1 of the VNA under magnetic field $\mu_0H_\mathrm{DC}=15\,\mathrm{mT}$ and angle $\theta=45^{\circ}$.
Red curves represent the fitting results using Eq.~(8).
The dip around $803\,\mathrm{MHz}$ (black dash line) in (c) corresponds to the SAW resonator mode.
(e)(f)~Color maps of evaluated coupling strength $g$ in (e) Device~1 and (f) Device~2 as a function of magnetic field $H_\mathrm{DC}$ and magnetic field angle $\theta$.
}
\label{fig:AMR+S11}
\end{center}
\end{figure}

Figure~\ref{fig:AMR+S11}(b) shows a magnetic field dependence of calculated resonance frequencies of the SW with the same wavenumber $k=2\pi/\lambda_\mathrm{SAW}=2\pi\times2.5\times 10^5 \,\mathrm{m^{-1}}$ as the SAW resonator mode in Ni film in Device~1 assuming no magnetic anisotropy~\cite{KS1986}. 
The green and red lines are the dependence at $\theta=90^{\circ}$ and $\theta=0^{\circ}$, while other angle cases lie between them. 
Figure~\ref{fig:AMR+S11}(b) implies that SW modes in Ni film are off-resonance with SAW resonator mode in the magnetic field region $10\,\mathrm{mT}-20\,\mathrm{mT}$ (blue background) where we perform experiments.

\subsubsection{Evaluation of coupled system parameters}
Here, we first summarize experimentally achieved parameters relevant to the SAW-SW coupled system shown in Fig.~\ref{fig:coupledsystem+setup}(a), such as the coupling rate $g$ and $\kappa_e$ appearing in Eqs.~(\ref{eq:Hme_quant}) and (\ref{eq:Hp_quant}), respectively, and the intrinsic dissipations represented by the rates $\kappa_p$ and $\kappa_m$ for the SAW resonator mode and the SW mode.
The experimental setup used in evaluating these parameters is shown in Fig.~\ref{fig:coupledsystem+setup}(b).
An itinerant microwave field generated from Port~1 of the VNA drives the system consisting of the SAW resonator and the SW mode via the IDT.

To evaluate the parameters $\omega_{p}$, $\kappa_{p}$, and $\kappa_{e}$ related to the SAW resonator, we measure the reflection coefficient $S_{11}$ from the SAW resonator in the condition where the SAW and SW are far-off-resonance ($\mu_0 H_\mathrm{DC}=200\,\mathrm{mT}$).
Under such condition, the SAW and the SW are not coupled, and the term including $g$ in Eq.~(\ref{eq:S11model}) can be neglected, and the reflection coefficient of the SAW resonator $S_{11}$ is simplified to
 \begin{equation}
     S_{11}(\omega)=\frac{i(\omega-\omega_p)-\frac{\kappa_p-\kappa_e}{2}}{i(\omega-\omega_p)-\frac{\kappa_p+\kappa_e}{2}}. \label{eq:S11SAW} 
 \end{equation}
By fitting the obtained result using Eq.~(\ref{eq:S11SAW}), we obtain the SAW resonator-related parameters as $\omega_p/2\pi=803\,\mathrm{MHz}$, $\kappa_p/2\pi=0.61\,\mathrm{MHz}$, and $\kappa_e/2\pi=7.81\,\mathrm{kHz}$.

When the magnetic field $\mu_0  H_\mathrm{DC}$ is reduced from $200\,\mathrm{mT}$ to $15\,\mathrm{mT}$, the $S_{11}$ signal will contain SW information shown in Eq.~(\ref{eq:S11model}) because the frequency difference between the SAW and SW will become smaller, as shown in Fig.~\ref{fig:AMR+S11}(b).
In other words, the term containing $g$ in Eq.~(\ref{eq:S11model}) cannot be ignored in this situation.
Figures~\ref{fig:AMR+S11}(c) and \ref{fig:AMR+S11}(d) show the electrical reflection spectra $S_{11}(\omega)$ of the SAW resonator for Device~1 at $\mu_0 H_\mathrm{DC}=15\,\mathrm{mT}$ and $\theta=45^\circ$.
From the fitting in Figs.~\ref{fig:AMR+S11}(c) and \ref{fig:AMR+S11}(d) based on Eq.~(\ref{eq:S11model}), we obtain $g/2\pi=5.67\,\mathrm{MHz}$, $\omega_m/2\pi=3.61\,\mathrm{GHz}$, and $\kappa_m/2\pi=1.07\,\mathrm{GHz}$.
Similarly, by fitting the $S_{11}$ spectra obtained at each magnetic field $H_\mathrm{DC}$ and field angle $\theta$, we obtain two-dimensional-color-maps of the evaluated coupling strength $g$ for Device~1 and Device~2, in Figs.~\ref{fig:AMR+S11}(e) and \ref{fig:AMR+S11}(f), respectively.
There is no apparent dependence on the magnetic field.
This feature is consistent with the fact that the coupling strength $g$ in Eq.~(\ref{eq:g_allparameter}) is independent of the magnetic field.
Note that as the magnetic field angle $\theta$ is varied, the parameters $\omega_{p}$, $\kappa_{p}$, and $\kappa_{e}$, related to the SAW resonator also vary due to magnetostriction (see Appendix B for details).
When fitting at an angle $\theta$, we use the parameters evaluated under far-off-resonance conditions at the $\theta$.

\begin{figure}[t]
\begin{center}
\includegraphics[width=8.6cm,angle=0]{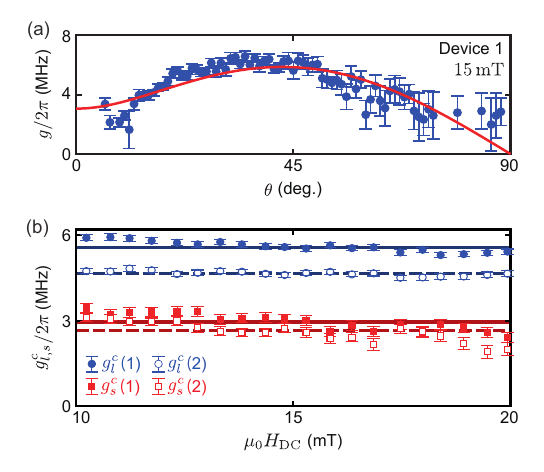}
\caption{
(a)~Cross section of Fig.~\ref{fig:AMR+S11}(e) at a magnetic field of $15\,\mathrm{mT}$ (indicated by white dash line) with fitting curve (red line) based on Eq.~(\ref{eq:g_forfit}). 
(b)~Magnetic field dependence of $g^c_l$ and $g^c_s$ for each device evaluated using Eq.~(\ref{eq:g_forfit}) from the results in Figs.~\ref{fig:AMR+S11}(e) and \ref{fig:AMR+S11}(f).
The four lines represent the average of the respective values.
}
\label{fig:gconst}
\end{center}
\end{figure}

Next, in preparation for estimating the practical constants $b_{1,2}D_{l,s}$, we decompose evaluated $g$ into $g^c_l$ and $g^c_s$ using Eq.~(\ref{eq:g_forfit}).
For instance, Fig.~\ref{fig:gconst}(a) shows a cross section of Fig.~\ref{fig:AMR+S11}(e) at the magnetic field $\mu_0H_\mathrm{DC}=15\,\mathrm{mT}$ (white dash line). 
From the fitting Fig.~\ref{fig:gconst}(a) based on Eq.~(\ref{eq:g_forfit}), we obtain 
$g^c_l/2\pi=5.4\,\mathrm{MHz}$ and
$g^c_s/2\pi=3.1\,\mathrm{MHz}$.
Figure~\ref{fig:gconst}(b) shows the magnetic field dependence of $g^c_l$ and $g^c_s$ obtained from the fitting of Figs.~\ref{fig:AMR+S11}(e) and \ref{fig:AMR+S11}(f) at each magnetic field.
The blue and red colors represent $g^c_l$ and $g^c_s$, and the pattern fill or not corresponds to Device~1 or Device~2.
The four lines represent the average of the respective values.
Each average value is noted in Table~\ref{table:SAW_parameters} and \ref{table:parameters_Device2}.

Here, to validate the evaluation method in this paper, we confirm the device geometry dependence of $g^c_l$ and $g^c_s$.
From Eqs.~(\ref{eq:g_longi}) and (\ref{eq:g_shear}), the ratio of $g^c_l$ and $g^c_s$ between each device depends only on $V_\mathrm{SAW}$ and $V_\mathrm{SW}$, and  can be written together as $R_{l,s}=g^{c}_{l,s}(1)/g^{c}_{l,s}(2)=\sqrt{\frac{V_\mathrm{SW}(1) V_\mathrm{SAW}(2)}{V_\mathrm{SW}(2) V_\mathrm{SAW}(1)}}$, where the number of the device is written in parentheses.
The reason why values other than $V_\mathrm{SAW}$ and $V_\mathrm{SW}$ can be regarded as common regardless of devices is that both devices are fabricated on the same substrate by the same process.
From the design values of $V_\mathrm{SW}(1,2)$ and the optically evaluated $V_\mathrm{SAW}(1,2)$ in Sec.~\ref{OI}, $R_{l,s}$ is obtained as $1.09$.
Whereas, from Fig.~\ref{fig:gconst}(b), the experimentally evaluated $R_{l}$ and $R_{s}$ are $1.20$ and $1.11$, respectively.
All values are close, confirming the validity of the method.

Finally, we estimate the practical constants $b_1D_l$ and $b_2D_s$, that depends only on the material and fabrication process.
Substituting the values of $g^c_l$ and $g^c_s$ for each device in Fig.~\ref{fig:gconst}(b) and the known values summarized in Table~\ref{table:SAW_parameters} into Eqs.~(\ref{eq:g_longi}) and (\ref{eq:g_shear}), we obtain $b_1D_l=8.1\times10^5\,\mathrm{J/m^3}$ and $b_2D_s=4.3\times10^5\,\mathrm{J/m^3}$ for Device~1 and $b_1D_l=7.4\times10^5\,\mathrm{J/m^3}$ and $b_2D_s=4.3\times10^5\,\mathrm{J/m^3}$ for Device~2.
Here, the saturation magnetization $M_{\rm{S}}$ in Table~\ref{table:SAW_parameters} is obtained by SQUID measurements.
Each constant is very well matched across devices.
This is evidence that our method can estimate the practical constant~$b_{1,2}D_{s,l}$ that characterizes the magnetoelastic coupling.

\begin{table}[t]
  \begin{center}
    \caption{Parameters of Device~1.} 
    \begin{tabular}{|c||c|} \hline
      Parameter & Value   \\ \hline \hline
      $g^c_l/2\pi$ & $5.58\pm0.02\,\mathrm{MHz}$  \\ \hline
      $g^c_s/2\pi$ & $2.95\pm0.04\,\mathrm{MHz}$  \\ \hline 
      SAW mode volume: $V_\mathrm{SAW}$  & $2.26\times10^3\,\mathrm{\mu m^3}$\\ \hline
      Ni film volume: $V_\mathrm{SW}$ &  $2.1\,\mathrm{\mu m^3}$ \\ \hline
      SAW velocity: $v_p=(\omega_{p}/2\pi)\lambda_\mathrm{SAW}$  & $3.2\times10^3\,\mathrm{m/s}$ \\ \hline
      SAW wavenumber: $k/2\pi$ &  $2.5\times10^7\,\mathrm{m^{-1}}$ \\ \hline
      Ni gyromagnetic ratio: $\gamma/2\pi$ &  $30.59\,\mathrm{GHz/T}$~\cite{MA1961}\\ \hline
      Ni saturation magnetization: $M_{\rm{S}}$ &  $5.25\times10^5\,\mathrm{A/m}$ \\ \hline
      $\rm{LiNbO_3}$ mass density: $\rho$ &  $8.9\times10^3\,\mathrm{kg/m^3}$~\cite{M1999} \\ \hline      
    \end{tabular}
  \label{table:SAW_parameters}
  \end{center}
\end{table}

\section{Discussion} \label{sec.discussion}
Our method demonstrated in this paper, which combines optical imaging and electrical measurements, is helpful for accurate understanding SAW-SW coupling systems.
Since SAW modes can be easily affected by the magnetic thin film on the substrate and by the structure and quality of the electrodes, it is essential to confirm the SAW mode and to evaluate the SAW mode volume by optical imaging.
For SAW resonators on a $\LiNb$ substrate, the upper limit of the SAW resonance frequency at which optical imaging is possible is about $1\,\mathrm{GHz}$.
However, in general, the resonance frequency of SW in magnetic thin films under an applied magnetic field that makes them a single magnetic domain is several gigahertz.
Hence, only a few materials were used in magnetoelastic coupling studies, and light was used sparingly.
The analysis presented in this paper, which can also be used under off-resonance conditions, allows the introduction of optical imaging and a more comprehensive selection of magnetic materials.
Our method will significantly contribute to the field, promising a more accurate evaluation of the magnetoelastic coupling.

Furthermore, our method provides an experimental way to evaluate the most practical constant $bD$ that characterizes the SAW-SW coupled system.
First, $b$ is the magnetoelastic coupling constant determined by the magnetic thin film's physical properties.
This value can vary depending on the substrate materials, surface conditions, and the fabrication processes.
Second, $D$ is the constant that determines the magnitude of the strain created by the SAW, which is determined by the substrate's physical properties.
This quantity is also sensitive to the magnetic thin film and metallic structures placed on top.
In other words, these constants $b$ and $D$ are challenging to estimate analytically.
On the other hand, their product $bD$ can be determined experimentally using our method and is directly comparable among studies.
Experimental evaluation and comparison of $bD$ is essential.

In this study, we assume that the SW is excited by the SAW and has the same wavenumber as the SAW.
Previous X-ray study~\cite{CS2020}, simultaneous detection of SAW and SW under on-resonance conditions in a setup similar to that in this paper, which verified the assumption.
Future studies with X-ray and light~\cite{KK2020,ZZ2021,GN2022,KK2024} will verify the above assumption under off-resonance conditions.

\section{Conclusion}
In conclusion, we demonstrated the novel method for evaluating the most practical constant $bD$ characterizing magnetoelastic SAW-SW coupling using electrical and optical measurements.
The constant $bD$ is directly comparable among studies.
The analysis that can be used even under off-resonance conditions is essential to the method's development.
The analysis enables the observation of the magnetoelastic couplings between SAW with resonance frequencies that are optically imageable and SW with resonance frequencies in the gigahertz range.
Our method will further advance the study of coupled SAW-SW systems.

\acknowledgments
This work was supported by JSPS KAKENHI (grant no. JP21K18145, JP22K14589, JP22KJ1995, JP23KJ1209,  JP23KJ1159, JP24H00007, JP24H02232), JST (grant no. JPMJFS2123, JPMJPR200A), MEXT Initiative to Establish Next-generation Novel Integrated Circuits Centers (X-NICS) (grant no. JPJ011438), and the Collaborative Research Program of the Institute for Chemical Research, Kyoto University.

\appendix
\section{Derivation of \\magnetoelastic coupling strength $g$}\label{Ap:g}
This Appendix describes how to obtain the theoretical expression of $g$ in Eq.~(11).

A magnetization vector $\boldsymbol{m}$ can be expressed as follows, considering the situation where the magnetization is saturated at an angle $\phi$.
\begin{equation}
     \boldsymbol{m}= \left(\begin{array}{c}m_x \\m_y \\m_z\end{array}\right)
     =\left(\begin{array}{c}M_{\rm{S}}\cos{\phi}-m_{\zeta}\sin{\phi} \\M_{\rm{S}}\sin{\phi}+m_{\zeta}\cos{\phi} \\m_{\xi}\end{array}\right).\label{eq:component_m}
\end{equation}
Where $m_{\xi}$ and $m_{\zeta}$ are the components perpendicular to the quantization axis, oriented in-plane and out-of-plane, respectively.
They are time-dependent.
Substituting Eq.~(\ref{eq:component_m}) into Eq.~(\ref{eq:Hme_classic2}) and extracting only the first-order term of the time-dependent magnetization, Eq.~(\ref{eq:Hme_classic2}) becomes
\begin{equation}
   H_\mathrm{me} = \int d^3r \left[b_1\,\frac{m_{\zeta}}{M_{\rm{S}}}\,\sin({2\phi})\,e_{xx} +  4b_2\,\frac{m_{\xi}}{M_{\rm{S}}}\,\cos({\phi})\,e_{zx}\right].\label{eq:Hme_classic3}
\end{equation}

Next, we represent the strain $e_{ij}$ using boson operators where $i,j\in\{x,y,z\}$.
Here, we consider plane waves located on the $xy$-plane and propagating in the $x$-axis direction.
Displacement and momentum density are written using the boson operators $a_{\mu k},a_{\mu k}^{\dagger}$ as follows~\cite{K1963,CD1999,GR2015},
\begin{equation}
    \begin{split}
       u_{i} =  &\sqrt{\frac{\hbar}{2\rho V_{\mathrm{SAW}}}} \sum_{k,\mu} \epsilon_{i\mu}(k) \omega_{\mu k}^{-1/2} f_{\mu k}(z) \\
       &\times (a_{\mu k}^{\dagger}e^{-ikx}+a_{\mu k}e^{ikx})\label{eq:ui}   
    \end{split}
\end{equation}
and
\begin{equation}
    \begin{split}
           \rho\dot{u_{i}} =&  \sqrt{\frac{\rho\hbar }{2V_{\mathrm{SAW}}}} \sum_{k,\mu} i \epsilon_{i\mu}(k) \omega_{\mu k}^{1/2} f_{\mu k}(z)\\
           &\times (a_{\mu k}^{\dagger}e^{ikx}-a_{\mu k}e^{-ikx}),\label{eq:momentum}
    \end{split}
\end{equation}
where $\mu=\{x,y,z\}$, $k$ is wavenumber, $\epsilon_{i\mu}(k)$ is unitary polarization vector, $\omega_{\mu k}$ is angular frequency of the acoustic waves, and
$f_{\mu k}(z)$ is dimensionless decay term that depends on the $z$-position and can be expressed as $f_{\mu k}(z)=\sum_{\mu}  A_{\mu}e^{-B_{\mu}kz}$.
Here $A_{\mu}$ and $B_{\mu}$ are dimensionless constants determined solely by the material constants.
From Eqs.~(\ref{eq:ui}) and (\ref{eq:momentum}), $e_{xx}$ and $e_{zx}$ can be quantized as follows,
\begin{equation}
    \begin{split}
       e_{xx} = & -ik \sqrt{\frac{\hbar }{2\rho V_{\mathrm{SAW}}}} \sum_{k} \omega_{k}^{-1/2}f_{xk}(z) \\
       &\times (a_{x k}^{\dagger}e^{-ikx}-a_{xk}e^{ikx})\label{eq:exx}    
    \end{split}
\end{equation}
and
\begin{equation}
    \begin{split}
        e_{zx} =&  \frac{1}{2} \sqrt{\frac{\hbar }{2\rho V_{\mathrm{SAW}}}} \sum_{k} \omega_{k}^{-1/2} \\
        &\times \{(a_{x k}^{\dagger}e^{-ikx}+a_{xk}e^{ikx})\frac{\partial f_{xk}(z)}{\partial z} \\ 
        &-ik(a_{z k}^{\dagger}e^{-ikx}-a_{zk}e^{ikx})f_{zk}(z) \}.\label{eq:ezx}
    \end{split}
\end{equation}
Note that the frequency $\omega_{\mu}$ is assumed to be constant.

Now consider the SAW resonator mode.
At the surface, i.e., $z=0$, the decay terms can be described as $f_{xk}(z)=D_1$, $\frac{\partial f_{xk}(z)}{\partial z}=kD_{2}$, and $f_{zk}(z)=D_3$ with the constants $D_1$, $D_2$, and $D_3$, which are determined only by the material properties.
For standing waves ($k = \pm k_0$), $e_{xx}$ and $e_{zx}$ are written as
\begin{equation}
    \begin{split}
           e_{xx} =&  2\sqrt{\frac{\hbar }{2\rho V_{\mathrm{SAW}} v_p k_0}} D_1 \cos({k_0x})(a_{x k_0}^{\dagger}+a_{xk_0})\label{eq:exx_standingwave}
    \end{split}
\end{equation}
and
\begin{equation}
    \begin{split}
       e_{zx} =&  \sqrt{\frac{\hbar }{2\rho V_{\mathrm{SAW}} v_p k_0}} \cos({k_0x})\\ 
       &\times \lbrace D_2(a_{x k_0}^{\dagger}+a_{xk_0})+D_3 (a_{z k_0}^{\dagger}+a_{zk_0}) \rbrace ,\label{eq:ezx_standingwave}        
    \end{split}
\end{equation}
where $\omega_{k_0}=\omega_{-k_0}=v_p k_0$.

Similarly, we also represent the micro-magnetization~$m_{k}$ using the boson operators.
Defining circularly polarised components as $m^{\pm}=m_{\zeta}\pm im_{\xi}$, $m^{\pm}$ is expressed using the magnon operators $c_k, c_k^{\dagger}$ as follows~\cite{GR2015},
\begin{equation}
   m^{+}=\sqrt{ \frac{\hbar \gamma M_{\rm{S}}}{2 V_{\rm{SW}}}}\sum_{k}e^{ikr}c_k.\label{eq:mplus}
\end{equation}
Using this expression, we get
\begin{equation}
   m_{\zeta}=\frac{1}{2}\sqrt{ \frac{\hbar \gamma M_{\rm{S}}}{2 V_{\rm{SW}}}}\sum_{k}\left( e^{ikr}c_k+e^{-ikr}c_k^{\dagger}\right)\label{eq:mzeta}
\end{equation}
and
\begin{equation}
   m_{\xi}=\frac{1}{2i}\sqrt{ \frac{\hbar \gamma M_{\rm{S}}}{2 V_{\rm{SW}}}}\sum_{k}\left( e^{ikr}c_k-e^{-ikr}c_k^{\dagger}\right).\label{eq:mxi}
\end{equation}

Substituting Eqs.~(\ref{eq:exx_standingwave}),~(\ref{eq:ezx_standingwave}),~(\ref{eq:mzeta}), and ~(\ref{eq:mxi}) into Eq.~(\ref{eq:Hme_classic3}), $H_\mathrm{me}$ is quantized as follows
\begin{equation}
    \begin{split} 
       H_\mathrm{me}=&\hbar \sqrt{\frac{\gamma k_0}{\rho v_p M_{\rm{S}}}} \frac{\int d^3r \cos^2{(k_0 r)}}
       {\sqrt{V_{\mathrm{SAW}}V_{\mathrm{SW}}}}\\
       &\times \lbrack b_1 \sin{2\phi} (c_{k_0}+c_{k_0}^{\dagger}) (a_{x k_0}+a_{x k_0}^{\dagger})\\
       &-2ib_2 \cos{\phi}  (c_{k_0}-c_{k_0}^{\dagger}) \\
       &\times \lbrace (a_{x k_0}+a_{x k_0}^{\dagger})D_2+(a_{z k_0}+a_{z k_0}^{\dagger})D_3 \rbrace \rbrack.
       \label{eq:Hme_quanta1}
    \end{split}
\end{equation}
To simplify Eq.~(\ref{eq:Hme_quanta1}), we further introduce a new operator,
\begin{equation}
    \begin{split}
           a_{tk}=&C (b_1 D_1 \sin({2\phi})\,c_{xk}-2ib_2 D_2 \cos({\phi})\,c_{xk}\\
           &-2ib_2 D_3 \cos({\phi})\, c_{zk} ), \label{eq:ct}
    \end{split}
\end{equation}
where $C=1/\sqrt{(b_1 D_1 \sin{2\phi})^2+\lbrace2b_2 (D_2+D_3) \cos{\phi}\rbrace^2}$ and $c_{tk}$ is boson operator ($t\in\{x,y,z\}$).
Applying the rotational wave approximation to Eq.~(\ref{eq:Hme_quanta1}), $H_\mathrm{me}$ can be written as
\begin{align}
       H_\mathrm{me}&=\hbar \sqrt{\frac{\gamma k_0 V_{\mathrm{SW}}}{\rho v_p M_{\rm{S}} V_{\mathrm{SAW}}} } \notag \\
       &\quad \times \sqrt{(b_1 D_l \sin{2\phi})^2+(2b_2 D_s \cos{\phi})^2} \\   
       &\quad \times(a_{tk_0}c_{k_0}^{\dagger}+a_{tk_0}^{\dagger}c_{k_0}) \notag \\
       &=\hbar g (a_{tk_0}c_{k_0}^{\dagger}+a_{tk_0}^{\dagger}c_{k_0}),
       \label{eq:Hme_quanta2}
\end{align}
where
\begin{equation}
   g = \sqrt{\frac{\gamma k_0 V_{\mathrm{SW}}}{\rho v_p M_{\rm{S}} V_{\mathrm{SAW}}} } \sqrt{(b_1 D_l \sin{2\phi})^2+(2b_2 D_s \cos{\phi})^2}.\label{eq:g_Appendix}
\end{equation}
Here, $D_l=D_1$, $D_s=D_2+D_3$, and the integral interval is the entire thin magnetic film.
Replacing $k_0=k$, $a_{tk_0}=\hat{a}$, $c_{k_0}=\hat{c}$, we obtain Eq.~(11).

\section{Magnetic field angle dependence of SAW resonator characteristics}\label{SM:SAWcaharacter_mag}

\begin{figure}[tb]
\begin{center}
\includegraphics[width=8.6cm,angle=0]{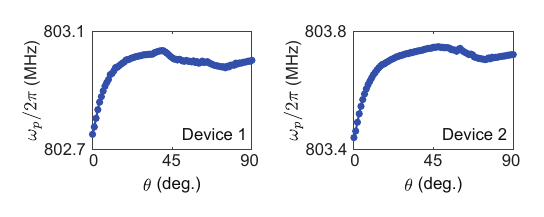}
\caption{Magnetic field angle dependence of the SAW resonance frequency $\omega_p$ in Device~1 and Device~2 at $\mu_0 {H}_\mathrm{DC} = 200\,\rm{mT}$.}
\label{fig:SAW200mT}
\end{center}
\end{figure}

This appendix describes the effect of the applied magnetic field on the SAW resonator through magnetostriction.
Figure~\ref{fig:SAW200mT} shows the magnetic field angle dependence of the resonance frequency $\omega_{p}$ of the SAW resonator modes for Device~1 and Device~2 at a magnetic field of $200\,\mathrm{mT}$.
The other parameters, $\kappa_{p}$ and $\kappa_{e}$, related the SAW resonator similarly depend on the magnetic field angle. 
These changes are due to the magnetostriction of the Ni thin film.

\begin{figure}[tb]
\begin{center}
\includegraphics[width=8.6cm,angle=0]{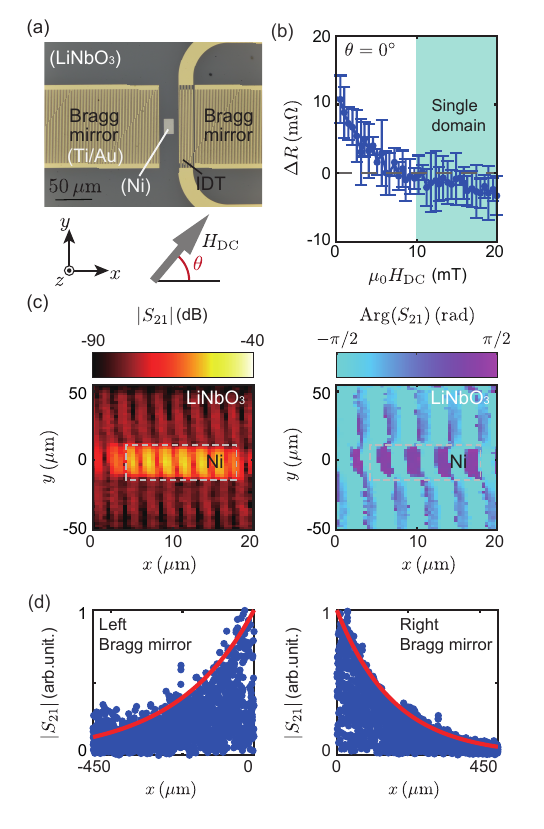}
\caption{(a) Microscope image of Device~2. (b) AMR measurement on Device~2 in $\theta = 0\,^{\circ}$. 
$\Delta R$ is obtained by subtracting the resistance for each magnetic field from the resistance at $\mu_0 H_{\mathrm{DC}}=200\,\mathrm{mT}$.
(c) Imaging results of the SAW resonator in Device~2. Color map of  amplitude and phase at a frequency $\omega/2\pi=803\,\mathrm{MHz}$. 
(d) Position dependence of optical signals in the Bragg mirror.}
\label{fig:Device2}
\end{center}
\end{figure}

\begin{table}[t]
  \begin{center}
    \caption{Parameters of Device~2.} 
    \begin{tabular}{|c||c|} \hline
      Parameter & Value   \\ \hline \hline
      $g^c_l/2\pi$ & $4.65\pm0.02\,\mathrm{MHz}$  \\ \hline 
      $g^c_s/2\pi$ & $2.66\pm0.02\,\mathrm{MHz}$  \\ \hline 
      SAW mode volume: $V_\mathrm{SAW}$  & $1.70\times10^3\,\mathrm{\mu m^3}$\\ \hline
      Ni film volume: $V_\mathrm{mag}$ &  $1.3\,\mathrm{\mu m^3}$ \\ \hline
    \end{tabular}
  \label{table:parameters_Device2}
  \end{center}
\end{table}

\section{Measurement results of Device~2}\label{SM:Device2}
This Appendix summarises the measurement results of Device~2.

Figure~\ref{fig:Device2}(a) shows an optical microscope image of Device~2. Device~1 and 2 use the same SAW resonator design.
The center position of the Ni thin film in Device~2 is the same as Device~1, but the $x$-axis length of the Ni thin film is $13\,\mathrm{\mu m}$.

Figure~\ref{fig:Device2}(b) shows the AMR measurement results when an external magnetic field is applied in the $x$ axis.
The result show that the resistance is saturated above $10\,\rm{mT}$, indicating that the Ni film in Device~2 is single magnetic domain.

Next, we show the optical measurement results of Device~2 and discuss the SAW mode and SAW mode volume.
Figure~\ref{fig:Device2}(c) shows two-dimensional plots of the amplitude and phase of the detected optical path modulation signal by scanning the optical spot in the $x$ axis every $0.5\,\mathrm{\mu m}$ and in the $y$ axis every $2\,\mathrm{\mu m}$ in a region ($20\,\mathrm{\mu m}\times 100\,\mathrm{\mu m}$) around center of the Ni film in Device~2.
The phase distribution in Fig.~6(c) shows that the SAW excited in the Ni film is a fundamental mode with a wavelength of $4\,\mathrm{\mu m}$ in Device~2 as in Device~1.

We next obtain the effective SAW mode volume in Device~2.
The difference from Device~1 is the length of the seepage $l$ into the Bragg mirrors.
Figure~6(d) shows that the SAW seeps about $220\,\mathrm{\mu m}$ ($160\,\mathrm{\mu m}$) into the left (right) Bragg mirror, namely, $l=48+220+160=422\,\mathrm{\mu m}$.

Finally, we estimate the constants $b_{1,2}D_{l,s}$ of Device~2.
Table~\ref{table:parameters_Device2} shows the parameters of Device~2.
Note that $v_p$, $k$, $\gamma$, $M_s$, and $\rho$ are the same as in Device~1 in Table~\ref{table:SAW_parameters}.
Substituting these parameters into Eqs.~(13) and (14), we obtain $b_1D_l=7.4\times10^5\,\mathrm{J/m^3}$ and $b_2D_s=4.3\times10^5\,\mathrm{J/m^3}$.


\end{document}